\documentclass[twocolumn,showpacs,preprintnumbers,amsmath,amssymb]{revtex4}
\usepackage{graphicx}               
\usepackage{dcolumn}                
\usepackage{bm}                     
\newcommand{\sech}{\mathrm{sech}}

\newcommand{\beq}{\begin{equation}}
\newcommand{\eeq}{\end{equation}}
\newcommand {\apgt} {\ {\raise-.5ex\hbox{$\buildrel>\over\sim$}}\ }
\newcommand {\aplt} {\ {\raise-.5ex\hbox{$\buildrel<\over\sim$}}\ }

\begin{document}

\title{Stratified shear flow instabilities at large Richardson numbers}

\author{Alexandros Alexakis}
\email{aalexakis@gmail.com}
\affiliation{Laboratoire de physique statistique,
Ecole Normale Superieure, rue Lhomond
75231 Paris, France}
\date{\today}

\begin{abstract}

Numerical simulations of stratified shear flow instabilities are performed
in two dimensions in the Boussinesq limit. 
The density variation length scale is chosen to be four times smaller
than the velocity variation length scale so that Holmboe
or Kelvin-Helmholtz unstable modes are present depending on the choice
of the global Richardson number $Ri$.
Three different values of $Ri$ were examined $Ri =0.2,\, 2,\, 20$.
The flows for the three examined values are all unstable 
due to different modes namely: the Kelvin-Helmholtz mode
for $Ri=0.2$, the first Holmboe mode for $Ri=2$, and the 
second Holmboe mode for $Ri=20$ that has been discovered recently
and it is the first time that it is examined in the non-linear stage.
It is found that the amplitude of the velocity perturbation of the second Holmboe mode 
at the non-linear stage is smaller but comparable to first Holmboe mode. 
The increase of
the potential energy however due to the second Holmboe modes is
greater than that of the first mode. 
The Kelvin-Helmholtz mode is larger by two orders of magnitude  
in kinetic energy than the Holmboe modes and about ten times
larger in potential energy than the Holmboe modes.
The results in this paper suggest that 
although mixing is suppressed at large Richardson numbers 
it is not negligible, and turbulent mixing processes
in strongly stratified environments can not be excluded.

\end{abstract}
\maketitle

\section{Introduction}
\label{Intro}

The destabilization of stratified layer due to the influence 
of a shear is a common phenomenon in nature. 
It occurs when the pressure gradients in the flow can overcome gravity
and overturn the fluid.
If the Reynolds number is large enough then this process quickly becomes turbulent,
diffusion is enhanced by the generation of small scales and the kinetic energy of the
flow is converted irreversibly to potential energy.
The rate kinetic energy is converted to potential energy and the total amount 
of potential energy gained  can be of crucial importance in determining the evolution
of many physically important systems, like 
the atmosphere \cite{Luce_2002,Gavrilov_2006}, 
 oceanic flows \cite{Farmer_1983,Armi_1988,Oguz_1990,Yoshida_1998}
and various astrophysical flows \cite{Rosner_2002,Alexakis_2004,Gauraud_2002}.  
In particular in this work the generation of turbulence and mixing
in strongly stratified environments is examined.
Such flows appear in accretion flows of Hydrogen and Helium 
into compact stellar object (white dwarf) composed of carbon and oxygen \cite{Rosner_2002,Alexakis_2004}  
A detailed analysis of the nuclear reactions of Hydrogen burning
has shown that energy release via catalytic nuclear reactions in which Carbon and Oxygen
play a crucial role can  lead to a nuclear runaway, in which a large fraction
of the accreted matter is expelled in the form of a shell; such a runway is
referred to in the astronomical literature as a nova.
However, how the needed Carbon and Oxygen of the compact star is 
mixed to the overlying accreted envelope in an environment where the acceleration
of gravity can six order of magnitude larger than terrestrial values is still an open question.
Turbulent mixing in the presence of strong stratification is also reported
in atmospheric and oceanic flows \cite{Galerpin_2007}.
It is of interests therefore to be able to estimate 
the amount of mass mixed
and how much increase of the potential energy can be generated
by shear flow instabilities as a function of the the control parameters of the system.
To that respect simple shear layer models of
velocity shear profiles $U(y)$ and density profiles $\rho(y)$ 
that depend only on the vertical coordinate of the system 
(here taken to be $y$)
have been proven a very useful in understanding the involved processes.

This work investigates a model first introduced by Hazel \cite{Hazel_1972}.
The model assumes a velocity profile given by
\beq
U_{_H}(y)=U_0 \tanh(y/L_{_U}) 
\label{BGU}
\eeq
and the density profile given by
\beq
\rho_{_H}(y)=\rho_{_0} [1 - \epsilon \tanh(y/ L_{\rho})].
\label{BGR}
\eeq
The ratio of the velocity variation length scale $L_{_U}$
to the density variation length scale $R=L_{_U}/L_{_\rho}$
is one of the control parameters of the system.
An other important control parameter in this system is
the Richardson number that expresses the ratio of the stabilizing
effect of gravity to the destabilizing effect of the shear.
For a general velocity and density profile the Richardson number
can be defined locally
at a height $y$ as: 
\[
Ri_{loc}(y) = -{g \frac{d\rho}{dy}}{\Big /}{\rho \left(\frac{dU}{dy}\right)^2  }
\]
where $g$ is the acceleration of gravity.
It also convenient to define a global Richardson number
in order to give a general measure of how strongly stratified 
the flow is.  In this work it is going to be defined as the local Richardson number at $y=0$,
that for the Hazel model becomes $Ri\equiv Ri_{loc}(0)= g\epsilon R/U_0^2L_{_U}$. 

Instabilities and generation of turbulence at large values Richardson numbers
seem somehow prohibited since the Miles-Howard theorem \cite{Howard_1961} guarantees 
that any flow in the inviscid non-diffusive
limit is linearly asymptotically stable if the local Richardson number is everywhere smaller than
1/4. This has restricted a lot of investigations to small values of global Richardson numbers.
However, depending on the details of the flow and the density stratification there 
can be regions in space where the local Richardson number can be smaller than 1/4 
while the global Richardson number is larger than 1/4.
These flows cannot be excluded from becoming unstable.

For the Hazel model
if $R$ is smaller than $\sqrt{2}$ the $Ri(y)$ has a unique
minimum at $y=0$ thus, linear instability  can exist only if the global 
Richardson number is smaller than 1/4. The unstable modes in this case are stationary 
(non-dispersive) waves concentrated around the $y=0$ plane where the flow has the strongest shear.
These modes are referred in the literature as Kelvin-Helmholtz modes (KH-modes)
due to their  resemblance with the instability of step function density and
velocity profile investigated by Lord Kelvin \cite{Kelvin_1910} and Helmholtz \cite{Helmholtz_1868}.
The linear stability of these modes has been investigated analytically 
by Miles \cite{Miles_1963} and numerically by Hazel \cite{Hazel_1972}.
The non-linear development of
Kelvin Helmholtz unstable modes has been investigated extensively in the
literature both experimentally \cite{Thorpe_1973,Thorpe_1968} and
numerically \cite{Caulfield_1994,Cortesi_1998,Caulfield_2000,Peltier_2003}. 
Their non-linear evolution leads to the formation of discrete billows
around the height of the strongest shear that curl the density gradients.
Secondary three dimensional instabilities at the nonlinear stage 
lead to the formation of a turbulent layer and fast mixing. 

For $R$ in the range $\sqrt{2}<R<2$ the local Richardson number $Ri(y)$ has 
two minima symmetrically placed around the origin with values
smaller than the global Richardson number. However in this range of $R$  
instability has been found only for values of the global Richardson number smaller 
than 1/4 with similar nonlinear evolution as with the $R<\sqrt{2}$ case.

For $R$ larger than 2 however $Ri(y)$ decays exponentially to 0 for large values of $y$. 
Thus, the Miles-Howard theorem can not guarantee linear stability of the flow
even for arbitrarily large values of the global Richardson number.
The parameter range $R>2$ therefor appears to be a good candidate for
instabilities in the strongly stratified limit. 
A numerical investigation of the inviscid linear instability problem in this 
parameter range was first performed by Hazel \cite{Hazel_1972} (and later on by
Smyth and Peltier \cite{Smyth_1989}). These early investigations
have shown that beside the Kelvin-Helmholtz instability that is 
confined to values of the global Richardson number smaller than 1/4,
new unstable modes are present.
These new modes appear as pairs of counter propagating
dispersive waves that are concentrated above and bellow the density ``interface". 
The unstable modes were found to be confined in
in a stripe (in the $Ri$ - wavenumber plane)
that extends to  arbitrarily large values of Ri. 
These results reproduce qualitatively the results
of a piece wise linear velocity and discontinuous density
profile introduced earlier by Holmboe \cite{Holmboe_1962}
and are referred in the literature as Holmboe modes.
Numerical investigations of the Holmboe instability 
have been performed in two \cite{Smyth_1988} and three
dimensions \cite{Smyth_1991,Sutherland_1994,Smyth_2003,Carpenter_2007}. 
It is worth noting that in reference
\cite{Smyth_2003} it was found that the Holmboe mode for $R=3$
resulted in larger increase of potential energy
than the Kelvin Helmholtz mode with $R=1$ although the latter one
had larger growth rate. 
Experimentally Holmboe modes have been investigated by 
by various groups 
 \cite{Browand73,Koop76,Lawrence_1991,Caulfield95,            Pawlak99,Pouliquen01,Zhu01,Hogg03,Negretti_2007}.
In the  nonlinear stage the unstable waves form of cusps, whose breaking
is responsible for mixing.

The persistence of the Holmboe instability at arbitrary large values of Ri makes them
better candidates for the generation of turbulence at strongly stratified environments. 
However, the growth rate of the unstable modes appears to decrease exponentially
with the Richardson number, and the presence of even small viscosity
restricts the region of instability to relatively small values of $Ri$. 

However, it was shown recently by the author \cite{Alexakis_2005,Alexakis_2007} that the unstable modes 
found by Hazel \cite{Hazel_1972} and Smyth \cite{Smyth_1989} are not the only ones present
in the Hazel model. 
When $R>2$ there is an infinite series of
unstable regions in the Richardson-wavenumber parameter space in the form of stripes.
Each new instability stripe appears at larger value of Ri
and corresponds to a different internal gravity
wave that becomes unstable when its phase speed becomes equal to the
maximum velocity of the flow \cite{Alexakis_2007}. These modes are going to be referred to
as higher Holmboe modes, and are going to be numbered with the order of appearance
as Ri is increased (first Holmboe mode, second Holmboe mode, \dots etc)
with the mode found by Hazel \cite{Hazel_1972} and Smyth and Peltier \cite{Smyth_1989}
being the first Holmboe mode.
It was further found in \cite{Alexakis_2005} that
for a fixed Ri the highest Holmboe mode, has the largest growth rate.
Therefor these newly discovered modes provide a new mechanism for 
generation of turbulence and mixing in strongly stratified flows.

The non-linear evolution of the higher Holmboe modes, has not been 
investigated in the non-linear regime neither numerically nor experimentally
since typical investigations so far have focused on small values
of the global Richardson number. The results of the linear theory
for the higher Holmboe modes are promising for the generation
of turbulence in strongly stratified environments, 
however turbulence and mixing can
only be addressed in the nonlinear stage of the evolution.

This work  examines the development of shear flow instabilities 
that span three orders of magnitude of the global Richardson Number.
The examined values $Ri=0.2$, $Ri=2.0$, $Ri=20$, correspond to three different
unstable modes: the Kelvin-Helmholtz mode for $Ri=0.2$,
                the first  Holmboe   mode for $Ri=2.0$
          and   the second Holmboe   mode for $Ri=20.$
Since this is the first numerical study of the second Holmboe mode
the investigation is restricted only to two dimensions, in order to 
get a basic understanding of the nonlinear development without 
the additional complications of secondary three dimensional instabilities. 
However, since properties of turbulence and mixing can be very different in 
three and two dimensions care is needed in the interpretation of the results,
and their implications to the physical systems. For this reason the results
in this paper will be restricted only to basic mechanisms involved
and the relative increase of kinetic and potential energy of the examined modes
without examining in detail mixing properties that would depend on the dimensionality 
of the system.

In the next section we introduce 
in detail the model that is going to be investigated,
examine the linear theory,
give the details of the numerical code,
and justify the choice of parameters.
Section III presents the results of the numerical simulations.
In the last section the results of this work are discussed
and conclusions are drawn.

\section{Methodology}


\subsection{The mathematical model}
\begin{figure}
\includegraphics[width=8cm]{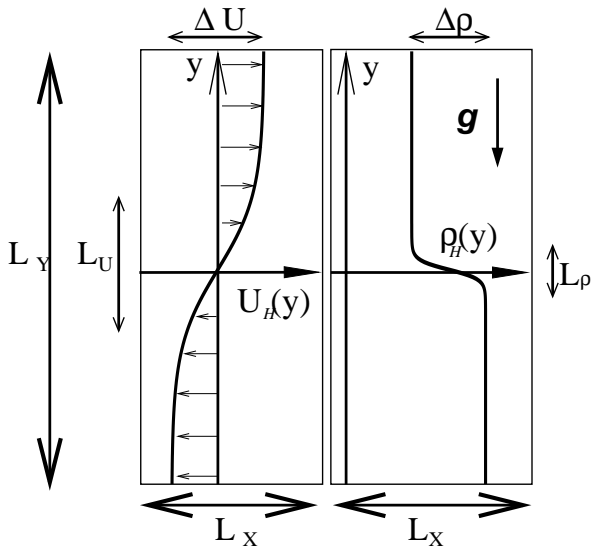}
\caption{\label{fig1} A sketch of the model under study}
\end{figure}

In this study a two dimensional incompressible flow  of an inhomogeneous in density fluid 
will be considered. The fluid is confined in a rectangular box of size $L_{_Y},L_{_X}$ with periodic
boundary conditions in the $x-$direction and free-slip ($u_y=0,\partial_yu_x=0$), 
no-flux ($\partial_y\rho=0$) conditions in the top and bottom boundary.
The Boussinesq equations for the evolution of the velocity and density field then read:
\beq
\partial_t {\bf u} + {\bf u \cdot \nabla u} = 
\frac{1}{\rho_{_0}} \nabla P -{\bf j} g \frac{\rho}{\rho_{_0}} +\nu \nabla^2 {\bf u} + {\bf i}F, 
\label{NS}
\eeq
\beq
\partial_t {\bf \rho} + {\bf u \cdot \nabla \rho} = \kappa \nabla^2 {\rho} + {S},
\label{AD}
\eeq
where ${\bf u}$ is the incompressible velocity field, and $\rho$
is the density field. The mean value of the density field is given by $\rho_{_0}$.
$\nu$ and $\kappa$ are the viscosity and the diffusivity of the fluid. $g$
is the acceleration of gravity assumed here to act in the negative $y-$direction.
$F(y)$ is a forcing function and $S(y)$ is density source/sink term 
with zero average so that the space averaged density $\langle\rho\rangle=\rho_0$ is conserved.
%
%
$F$ and $S$ are chosen so that 
$F=-\nu \partial_y^2 U_{_H}$ and $S=-\kappa \partial_y^2 \rho_{_H}$ where $U_{_H}(y)$ and $\rho_{_H}(y)$
are given in eq. \ref{BGU},\ref{BGR} with $y=0$ corresponding to the mid plane of our box.  
With this choice ${\bf u=i}U_{_H}(y)$ and $\rho=\rho_{_H}(y)$ are exact solutions of 
the Boussinesq equations \ref{NS},\ref{AD}.
(To be more exact we need to add an exponentially small term in equations \ref{BGU},\ref{BGR}
(proportional to $ -2U_0y/L_y \sech^2(L_{_Y}/2L_{_U})$, and $ -2\rho_0y/L_{_Y} \sech^2(RL_{_Y}/2L_{_U})$) in order for the
laminar solutions $U_{_H}(y)$ and $\rho_{_H}(y)$ to satisfy the boundary conditions. This term was included in the numerical simulations
for consistency although at the examined box sizes presented here it didn't seem to play an important role,
however at smaller box sizes (not presented here) it helped to avoid the formation of boundary layers
at $y=\pm L_{_Y}/2$.)
A sketch of the model that is investigated is shown in figure \ref{fig1}. 

To non-dimensionalise the system we are going to use the velocity amplitude $U_{_0}$, the velocity length scale
$L_{_U}$ and the density $\rho_0$. Thus, in the results presented in the next section
all length scales are in units of $L_{_U}$, time scales in units of $L_{_U}/U_0$ and energy
in units of $\rho_0U_0^2$. 
This leads to 4 non dimensional control parameters that control the Hazel model, namely:
the Richardson number $Ri$ defined as $Ri \equiv Ri_{loc}(0)= g \epsilon L_{_U}^2/U_0^2 L_\rho$,
the Reynolds Number $Re \equiv U_{_0}L_{_U}/\nu$,
the Prandtl (or Schmidt) number $Pr \equiv \nu/\kappa$ and 
the ratio of the velocity length scale to the density length scale $R=L_{_U}/L_{_\rho}$.
In addition to the just mentioned parameters of the Hazel model
there are two more parameters in the examined system due to the finite size of the computational box
$\ell_{_Y} \equiv L_{_Y}/L_{_U}$ and $\ell_{_X} \equiv L_{_X}/L_{_U}$.



\subsection{The linear instability problem} 

Before investigating the nonlinear problem we need to examine the 
linear stability problem for the diffusive and dissipative system 
in parameter range that is going to be examined with the numerical 
simulations. 
The linear stability theory considers the evolution of infinitesimal 
perturbations to the background density and velocity profiles.
The velocity perturbation is going to be written in terms of a stream function $\psi$ as
${\bf u} - {\bf i} U_{_H}(y)= {\bf i} \partial_y \psi - {\bf j} \partial_x \psi$
and the density perturbation as $\rho-\rho_{_H} = \theta$.
A normal mode expansion will be assumed
\[\psi=\sum \tilde{\psi}_ke^{ik(x-ct)},\,\,\,\,\theta=\sum \tilde{\theta}_ke^{ik(x-ct)}\]
with $k$ being the wave number in the $x$-direction and $c$ the complex phase speed.
If the imaginary part of $c$ is greater than zero then the normal mode will grow 
exponentially with growth rate $\gamma= k \Im(c)$.
Linearizing equations \ref{NS},\ref{AD} with respect to $\psi$ and $\theta$ 
lead to the eigen value problem
\begin{eqnarray}
\left[ (U_{_H}-c) D^2 - U_{_H}'' -\frac{\nu   }{ik} (D^2)^2 \right] {\tilde{\psi}  }_k + \left[\frac{g}{\rho_{_0}}\right]{\tilde{\theta}}_k =0 \nonumber \\
\left[ (U_{_H}-c)                -\frac{\kappa}{ik}  D^2    \right] {\tilde{\theta}}_k + \rho_{_H}'{\tilde{\psi}}_k =0
\label{LT}
\end{eqnarray}
for the eigenvalue $c$. Here, prime indicates differentiation with respect to the $y$-coordinate and 
$D^2$ stands for the operator $D^2=\partial_y^2-k^2$. 
If we assume zero viscosity and diffusivity equations \ref{LT} 
simplify to the Taylor-Goldstein equation (see for example \cite{Drazin_book}).

In this work however the effect of viscosity and diffusivity can not be neglected
and the full eigen-value problem \ref{LT} needs to be examined.
The eigen value problem was solved numerically by expanding the two perturbative fields $\psi_k,\theta_k$
in a finite sum of sines (stream function) and cosines (density).
With this choice the two fields always satisfy the boundary conditions. 
The eigen-value problem \ref{LT} then becomes a matrix eigen-value problem
of the form $A x =c Bx$ which is then solved using the linear algebra package {\sc lapac}.
For  Re=500 and R=4 that will be our choice in the numerical simulations, 
128 modes were sufficient for
the code to converge to the third digit of the growth rate for 
all the modes except the ones close to the stability boundaries. 
For the modes with the real part of $c$ close to $\pm$1 
(that correspond to the small wave number boundary of Holmboe instability)
the eigen-value code had problem converging with such accuracy. 
The results in this paper are restricted 
modes with $|c|<0.98$ and as a result the stability boundaries presented here
extend slightly more to the left.

\begin{figure}
\includegraphics[width=8cm]{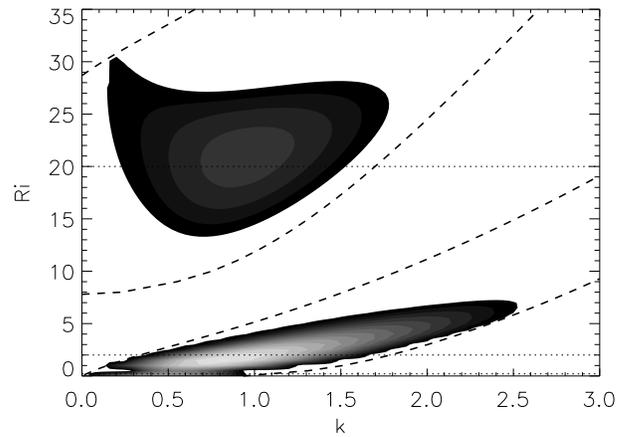}
\caption{\label{fig2} The stability diagram in the Richardson wavenumber plane.
shaded regions show the the locations of positive growth rate 
(light regions correspond to high growth rate), White regions are stable.
The dashed lines indicate the stability boundaries of the inviscid, non-diffusive system. 
The dotted horizontal lines indicate the three examined Richardson numbers.}
\end{figure}

Figure \ref{fig2} displays the stability diagram in the Richardson-wavenumber plane for 
$Re=500$, $R=4$, $Pr=1$ and $\ell_{_Y}=8\pi$.  
Shaded regions show the the locations of positive growth rate
(light regions correspond to high growth rate).
Different regions of instability can be seen.
The shaded region for large Richardson numbers ( $12\aplt Ri \aplt 30$) 
corresponds to the second Holmboe mode, while the shaded region for smaller values of $Ri\aplt 8$
corresponds to the first Holmboe mode. The Kelvin Helmholtz modes are restricted to small
values of $Ri<0.25$ and can not be seen clearly in this diagram. 
The Kelvin-Helmholtz instability region 
appears as a thin shaded region
close to $Ri=0$ in the range $0<k<1$.  
The dashed lines indicate the stability boundaries of the inviscid, non-diffusive system.
It can be clearly seen that even a small viscosity ($Re=500$) has significantly reduced
the unstable regions. 

\begin{figure}
\includegraphics[width=8cm]{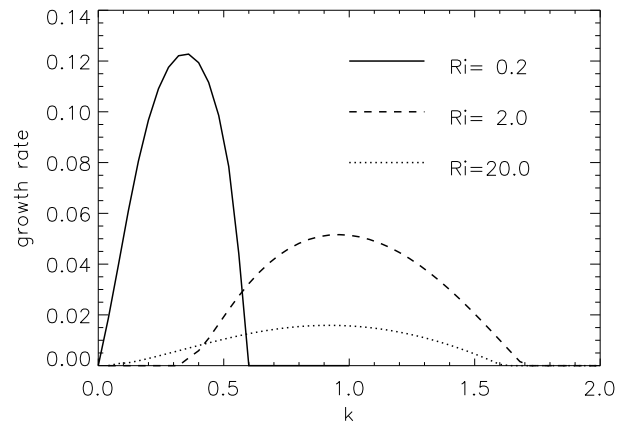}
\caption{\label{fig3} The growth rate for the three examined values
of the Richardson number: $Ri=0.2$ Kelvin-Helmholtz Solid line
$Ri=2.0$ first Holmboe mode dashed line, $Ri=20$, second Holmboe mode.}
\end{figure}

Figure \ref{fig3} shows the growth rates at the same parameter regime as a function of the wave number 
for the three examined values of the Richardson number $Ri$.
As can be seen the Kelvin-Helmholtz mode has the largest growth rate (solid line)
                                                             $\gamma\simeq 0.127$ for $k\simeq  0.36$, 
the first Holmboe mode (dashed line) has maximum growth rate 
                                                             $\gamma\simeq 0.0516$ for $k\simeq 0.96$,
and finally the second Holmboe mode has maximum growth rate 
                                                             $\gamma\simeq 0.0159$ for $k\simeq 0.92$.
It is worth noting when comparing the Kelvin Helmholtz modes with the second Holmboe mode that
although the Richardson number has changed by a factor of 100
the growth rate has been decreased by a factor less than 10.

Different values of $Re$ have also been examined.
Here, it is just noted that unstable second Holmboe modes
were found for Reynolds numbers $Re\apgt 160$.
The maximum growth rate for the second Holmboe mode 
at $Ri\simeq22,\,k\simeq 1$ becomes roughly 
Reynolds number independent when $Re\apgt 500$
(a difference less than $5\%$ was noted for the
growth rate for $Re=500$ and $Re=1000$).


\subsection{The numerical method}

To solve equations \ref{NS}, \ref{AD} a psedospectral code was used.
The velocity field was written in terms a stream function $\psi$ as
${\bf u}= {\bf i} \partial_y \psi - {\bf j} \partial_x \psi$.
The stream function and the density field were expanded in sines and
cosine modes (respectably) in the $y-$direction and in Fourier modes 
in the $x-$direction. 
\[\psi=\sum_{\bf k} \tilde{\psi}_{\bf k}e^{ik_xx}\sin(k_yy') ,\,\,\rho=\sum_{\bf k} \tilde{\rho}_{\bf k}e^{ik_xx}\cos(k_yy')\]
where $y'=y-\ell_{_Y}/2$ and $k_x=2\pi n/\ell_{_X}$, $k_y=\pi m/\ell_{_Y}$ with $n,m$ integers. 
The spatial derivatives were calculated in 
sine-Fourier space while products of fields were calculated in real space.
Dialiasing was achieved using the 2/3 rule.
The fields were  advanced in time using a third order Runge-Kuta method.

The adopted resolution for each performed run was decided  based on the spectral properties
of the two fields. A run was considered well resolved if the gradients
of the two advected quantities density $\rho$ and vorticity $w=\nabla \times {\bf u}$ were 
sufficiently resolved so that the peek of the spectrum of $\nabla \rho $ and $\nabla w$ 
was much larger than its value at the largest wave number.
 
Special care needs to be taken to determine the time step that satisfies the 
Courant, Friedrichs, Lewy 
(CFL) criterion \cite{CFL}.
The time step $\Delta t$  used in the code should be smaller than the grid size $\Delta x$ divided by the maximum speed in the problem. 
There two relevant speeds in the examined problem one given by the flow velocity and one given by the gravity wave speed.
For sufficiently sharp density interfaces the gravity phase speed scales like $c\sim \sqrt{g/k}$.   
Typically in simulations of the Kelvin-Helmholtz or the first Holmboe mode the criterion 
$\Delta t > \Delta x /U_{_0}$ is sufficient to satisfy CFL since $U_{_0}$ is the largest speed in the system
(Note that the phase speed of the unstable Holmboe modes has to be with in the range of $U$ for the instability to exist). 
An exception to this rule is the case where boxes much larger ($L_x \gg L_{_U}$) 
than the typical unstable wave length are considered.
In this case it is the phase speed of the longest gravity wave determines the allowed time step.
 
When investigating the second Holmboe mode one needs to be more careful because 
the unstable gravity wave mode is not the fastest mode in the system but rather the first Holmboe mode which is stable
and has phase speed much larger than $U_0$. As a result a much smaller time step was used the for the simulations
of the second Holmboe mode. In practice the time step used was based on the formula $\Delta t =a \Delta x/ \max(U_0,\sqrt{gL_{_X}/2\pi})$
for $a=0.1$.    
  

\subsection{Parameter choice}

In principle it is desirable to study this system in the limit
$Re,\ell_{_X},\ell_{_Y}\to\infty$ in order to make contact with
flows that appear in nature. 
In addition large numbers of $Ri\gg1$ should be considered for the
astrophysical flows that were mentioned in the introduction.
However, computational constraints put strong restrictions on the 
parameter space that can be examined. 

$R$ was fixed to the value $R=4$ that is sufficiently larger than the critical value
$R=2$ but not too large so that the density interface is not to sharp to resolve.
The three examined values of the Richardson number $Ri=0.2,\,2,\,20$ were based on the results of the
linear theory. Both for the first and the second Holmboe mode the values of $Ri=2$ and $Ri=20$
are very close to the value for which the growth-rate obtains its maximum value.
For the Kelvin Helmholtz mode we could have chosen a value of $Ri$ that is arbitrarily small, 
since the maximum growth rate is obtained for $Ri=0$. The choice of $Ri=0.2$ was made
so that the three values are different by an order of magnitude each.

The choice for $\ell_x$ was based on resolution and time step restrictions.
As mentioned in the previous section increasing $\ell_x$ not only increases
the number of modes that are needed for a fixed resolution but also decreases the
the time step. 
Two choices were made for this parameter. 
First single mode perturbations were examined and
$\ell_x$ was fixed to the value $\ell_x=2\pi$ for the Holmboe modes
so that the evolution of the most  
unstable wavenumber $k\simeq 1$ is captured. Similarly for the
Kelvin Helmholtz instability the choice of $\ell_x=8\pi$ was made.
For a second set of runs that more than single wavelengths was perturbed
the choice $\ell_x=8\pi$ was made for all three values
of the examined Richardson numbers so that also sub-harmonic coupling can also be captured.
$\ell_y$ has to be sufficiently large so that the boundaries play minimal role in the
the development of the instability. In the simulations the value $\ell_{_Y}=8\pi$ was chosen.
It proved to be sufficient for all modes. 

The Reynolds number was based on resolution requirements. 
In the examined cases a uniform grid with 512 grid points in the $y$ direction proved to be 
sufficient to resolve flows with $Re=500$.
Larger runs of 1028 grid points in the $y$-direction with the same Reynolds number were also performed but for 
shorter times that were in very good agreement with the 512 runs.

The choice of the Prandtl (Schmidt) number $Pr=1$ was also based on resolution requirements.
In most cases that Holmboe instability appears are with $Pr$ much larger than one. However such a 
choice would require an even finer grid to resolve the resulting thin filaments of density gradients. 
In \cite{Carpenter_2007} different grid sizes were considered for the evolution of the velocity and the density field
so that large Prandtl can be considered efficiently. Such an option will be considered in future investigations.

At this point, the choice of including the density source and forcing functions $S,F$ should also be justified.
For sufficiently large $Re$,$RePr$ the evolution of the background
fields due to viscosity and diffusion happens in a much longer time scale 
than the time scale given by the growth rate of the instabilities. 
In such case the evolution of the unstable modes is not expected to be affected 
significantly by the slow evolution of the background fields and the inclusion of $S,F$
would not be necessary.  
However since the second Holmboe mode has a relatively small growth rate very large Reynolds numbers 
need to be considered for such an assumption to hold.
In this work the inclusion of the forcing functions makes the 
the background fields exact solutions of the the Boussinesq equations \ref{NS},\ref{AD} and allows the investigation of
the instability problem free from long time scale approximations.
%

\section{Results}

\subsection{Single mode perturbations}

First the  evolution of a single unstable mode is examined.
The size of the domain $\ell_{_X}$ is chosen so that 
it is close to the wavelength of the fastest growing mode.
For the Kelvin-Helmholtz mode ($Ri=0.2$) a box of size $\ell_{_X}=8\pi$ was chosen 
and for the two Holmboe modes 
a box size of $\ell_x=2\pi$.

The initial conditions for the runs consisted of the background fields given in eq. \ref{BGU} and \ref{BGR} 
plus a small perturbation in the velocity. A perturbation in the density field was not added because
the energy of such perturbation will depend on the value of the gravitational acceleration $g$ that is essentially
varied with the Richardson number in the three examined cases. 
The form of the perturbation in terms of the stream function is 
\begin{equation}
\psi_{pert} = f_s(y) sin(kx +\phi_r) + f_{as}(y)cos(kx + \phi_r) 
\label{pert}
\end{equation}
where $\phi_r$ is a random phase, $k$ is the smallest wavenumber in the examined box and $f_s(y)$,$f_{as}(y)$ 
are a symmetric and an antisymmetric exponentially decreasing functions  
that satisfy the boundary conditions. The form of the perturbation was chosen so that there 
is no a priori exclusion of symmetric or antisymmetric solutions.

\begin{figure}
\includegraphics[width=8cm]{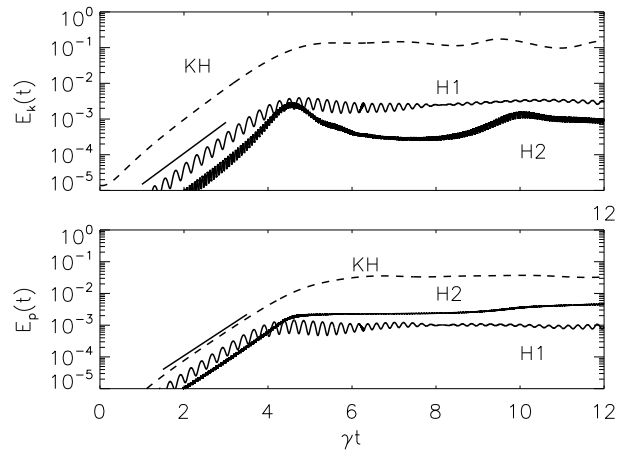}
\caption{\label{fig4} Top panel: Time evolution of the kinetic energy of the perturbation
of  the Kelvin-Helmholtz mode dashed line (KH), 
    the first Holmboe mode oscillating line (H1), 
and the H2 mode fast oscillating line (appears like thick line) (H2).}
\end{figure}

After short transient time the unstable eigenmode becomes dominant and an 
exponential increase of the space averaged kinetic energy of the perturbation
\beq
E_k(t)=\frac{1}{  \ell_{_X} \ell_{_Y}  } \int \frac{1}{2} \rho_0 ({\bf u-i}U_{_H})^2  dxdy
\label{kinetic}
\eeq
and the space averaged potential energy of the perturbation
\beq
E_p(t)=\frac{1}{\ell_{_X}\ell_{_Y} } \int (\rho-\rho_{_H}) g \rho y dxdy 
\label{potential}
\eeq
is observed.
Figure \ref{fig4} shows the time evolution of the kinetic energy $E_k$ (top panel) and
the potential energy $E_p$ (bottom panel) of the perturbation for the three examined values of $Ri$
in a log-linear plot.
Note that since the definition of the potential energy is linear with respect to $\rho$ the increase of 
the potential energy of the perturbation $E_p$ is equal to the increase of the potential energy of the whole system.
The evolution of the energy of the Kelvin-Helmholtz mode for $Ri=0.2$ is plotted with a dashed line and is marked as KH.
The oscillating line marked as H1, corresponds to the energy of the first Holmboe mode with $Ri=2.0$.
The fast oscillating line (that appears as a thick line) marked as H2, corresponds to the second Holmboe mode for $Ri=20$.
The time axis has been rescaled using the growth rate $\gamma$
of each mode, (obtained from the linear theory)
in order to fit the three lines in one plot for comparison.
As a result in the linear stage the three modes appear to grow with the same rate.
The straight lines (at $1<\gamma t<3$) show the prediction of the linear theory $e^{2\gamma t}$.
All three modes were started with the same perturbation energy, however the Kelvin-Helmholtz mode
had a shorter transient time and started growing sooner than the Holmboe modes and for this reason it appears
as if the KH mode starts with more energy.
 
In the linear stage of evolution the energy of Kelvin-Helmholtz grows like a pure exponential $\sim e^{2\gamma t}$.
Unlike the Kelvin-Helmholtz case that a single stationary mode is present
the Holmboe modes appear in pairs of opposite traveling waves.
The observed energy evolution of the Holmboe modes shown in figure \ref{fig4} is the energy of 
the sum of the two waves (left and right traveling waves) each one of which grows like $\sim e{(\gamma \pm i\omega )t}$.
As a result the total energy scales like $\sim e^{2\gamma t}[1+\alpha cos(2\omega t)]$
where $\alpha$ is a constant smaller than one that is proportional to the overlapping of the the eigenfunctions of the 
two waves.
This leads to the observed oscillations of the kinetic and potential energy in figure \ref{fig4}.
The frequencies of oscillations for the first and the second Holmboe mode are similar but in figure \ref{fig4} 
since the time axis has been rescaled with the growth rate the oscillations of the second Holmboe
mode appear as of higher frequency. It is also worth noting that the amplitude of the oscillations of the second Holmboe mode
are smaller than those for the first Holmboe mode implying that in the former case there is less overlapping of the two waves. 

For $\gamma t$ larger than 4, the nonlinearities become important and the exponential growth stops.
The amplitude of the energy however that this transition occurs is very different for the three modes.
In the non-linear stage the amplitude of the kinetic energy of the KH mode is roughly two orders of
magnitude bigger than that of the two Holmboe modes. 
The exponential growth of the first Holmboe 
and the second Holmboe mode stops at the same amplitude ($E_k\sim10^{-3}$) but the evolution of the 
kinetic energy the second Holmboe mode is followed by a decrease in amplitude  and then a
subsequent rise at later times.
This process appears to operate in much longer time scale than the wave crossing frequency
and is possibly related to the weak coupling of two counter propagating waves. 
The kinetic energy of the second Holmboe mode however always appears to be smaller that the kinetic energy of the 
first Holmboe mode roughly by a factor of three.

The behavior of the potential energy in the nonlinear stage has a different behavior.
The potential energy increase of the second Holmboe mode appears to exceed the potential energy
of the first Holmboe mode.
Although it is still smaller than the potential energy of the Kelvin Helmholtz
mode the difference is much smaller than that of the kinetic energy.
It is worth noting that 
apart from the fast oscillations, on longer time scales, the potential energy
is increasing monotonically this suggests that
the potential energy that has been gained has been irreversibly mixed
so that it cannot be returned to the flow. 
 
\begin{figure}
\includegraphics[width=8cm]{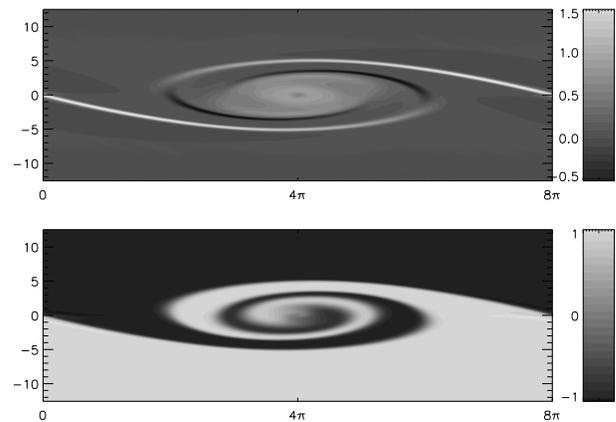}
\caption{\label{fig5} Shadow-graphs of the vorticity (top panel) and the
density (lower panel) field at the non-linear stage for the Kelvin Helmholtz instability.}
\end{figure}

The structures that develop at the nonlinear stage of evolution of 
the Kelvin Helmholtz and the first Holmboe mode have been studied before in the literature. 
Here the results of these runs are also presented for comparison 
with the second Holmboe mode that is examined here for the first time.
Figure \ref{fig5} shows the resulting structures of the Kelvin Helmholtz instability at
the nonlinear stage. The top panel shows a shadow-graph of the vorticity and the bottom 
panel shows a shadow graph of the density stratification. 
The Kelvin- Helmholtz mode has  lead to the well observed pattern
where the vorticity and the density layer have rolled up.
It is noted here that the snapshot that was taken at $\gamma t\simeq 6$ has already past the stage 
that secondary three dimensional instabilities are expected to appear if the study was in three dimensions.

\begin{figure}
\includegraphics[width=8cm]{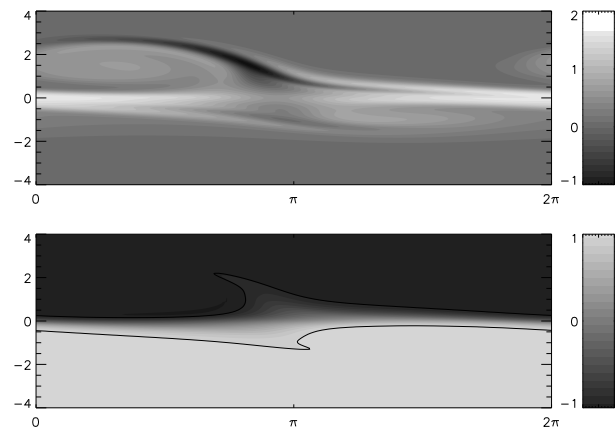}
\caption{\label{fig6} Shadow-graphs of the vorticity (top panel) and the
density (lower panel) field at the non-linear stage for the first Holmboe mode instability.}
\end{figure}

Contrary to the Kelvin-Helmholtz instability the first Holmboe instability leads to 
a pair of gravity waves coupled with two vortices above and below the density interface that travel in opposite 
directions. The gravity waves form cusps and eject material and thus mix the 
heavy fluid below the interface with the lighter fluid on top.
A snapshot of the vorticity (top panel) and density (lower panel) fields are shown in \ref{fig6}.
The solid black lines shown in the shadow-graph of the density are the contour lines 
that indicate the levels that the variation of density ($\rho-\rho_0$) 
has $95\%$ of its maximum (bottom line) and minimum (top line)
value. 

\begin{figure}
\includegraphics[width=8cm]{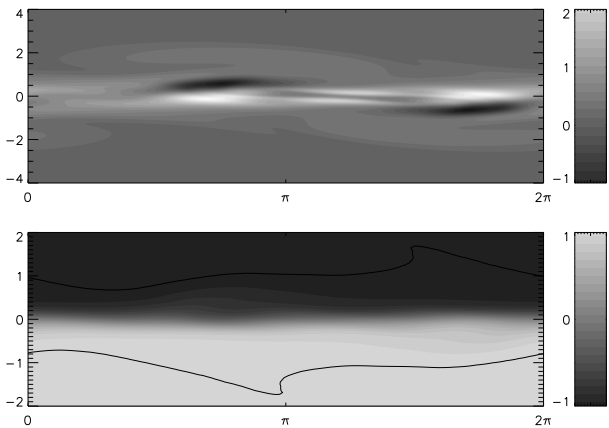}
\caption{\label{fig7} Shadow-graphs of the vorticity (top panel) and the
density (lower panel) field at the non-linear stage for the second Holmboe instability.}
\end{figure}

Finally shadow-graphs of the vorticity (top panel) and density (bottom panel)
of the second Holmboe mode are shown in figure \ref{fig7}.
The second Holmboe mode also consists of two counter propagating gravity waves. 
The gravity waves form cusps
and mixing is the result of the breaking of these cusps just like the mechanism
for mixing of the first Holmboe mode.
However when comparing the structures of the first and the second Holmboe mode
there are some differences that should be noted.
First it can be seen that the the vorticity field has a more complex structure
for the second Holmboe mode that involves the coupling of two pairs of counter rotating vortices 
close to the density interface.
The density structures are more similar for the two modes however the second  Holmboe mode
the cusps that form in the gravity waves are much weaker and appear at higher levels of 
density. The density contour lines that are drawn in the lower panel of  figures 
\ref{fig6},\ref{fig7} indicate the levels that the variation of density ($\rho-\rho_0$) 
has $99\%$ of its maximum (bottom line) and minimum (top line) value
(unlike the contour lines in figure \ref{fig4} for the first Holmboe mode where 
levels of $95\%$ was shown).
This implies that for the second Holmboe mode the mixing events that are related with breaking of the cusps
happen at larger heights where the density gradients are weaker and thus the mixing rate is slower.

\subsection{Multi-mode perturbation}
The situation of a single mode perturbation is somehow idealistic,
in more realistic situations more than one wavelength will become unstable 
and their coupling at the nonlinear stage could affect the resulting mixing rates.
Furthermore in the previously discussed set of runs different box sizes $\ell_{_X}$ 
were used. For a more fair comparison we need same box sizes and a more general
perturbation than the excitation of just a single wavelength.
For this reason a second set of runs was examined for the same Richardson numbers
as in the previous subsection. Here the box width $\ell_{_X}$ was set to $\ell_{_X}=8\pi$ for all
runs. The initial perturbation in the stream function that was introduced consisted of 
a sum 
of perturbations of the form of eq. \ref{pert} for ten wave numbers $k=2\pi n/\ell_{_X}$ (for $n=1,\dots,10$).
The density field was left again unperturbed.

\begin{figure}
\includegraphics[width=8cm]{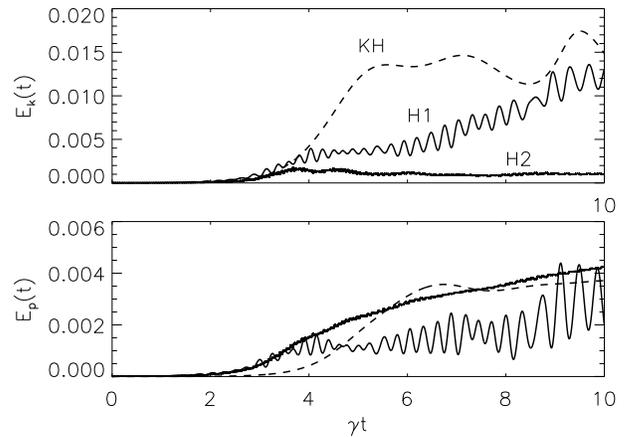}
\caption{\label{fig8} The evolution of the kinetic energy (top panel) and
potential energy of the perturbation for the multi-mode runs.
The kinetic energy of the Kelvin Helmholtz mode here shown with a dashed line
and marked by KH has been {\underline{decreased by a factor of 100}}
the potential energy has been {\underline{decreased by a factor of 10}}.
The oscillating solid line marked marked by H1 corresponds to the first Holmboe mode.
The solid line marked by H2 corresponds to the second Holmboe mode. }
\end{figure}

The evolution of the kinetic (top panel) and potential energy (lower panel) of the perturbation
is shown in figure \ref{fig8}. Unlike figure \ref{fig4}
a linear scaling for the $y$-axis is used so that the nonlinear stage is more clearly displayed.
 However because the energy of the Kelvin Helmholtz mode is much larger than the Holmboe modes
it has been rescaled by dividing the kinetic energy by a factor of 100 and the potential energy by a factor of 10.
As in figure \ref{fig4} the time scale has been rescaled with the growth rate of each mode.


%

The kinetic and potential energy of the Kelvin Helmholtz perturbation has little difference
from the previously examined case and obtains similar values of kinetic and potential energy
in the nonlinear stage as with the run examined in the previous section.
This is because both runs were performed in a similar size box with only the initial perturbation
being changed. Only small changes were observed in the resulting structures
of the vorticity and density field from the single mode run and are not shown here.

\begin{figure}
\includegraphics[width=8cm]{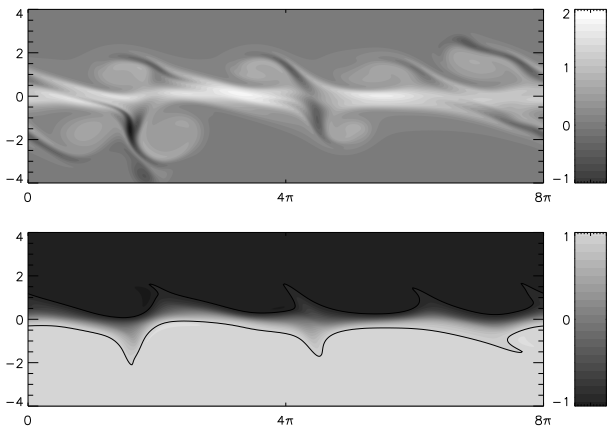}
\caption{\label{fig9} Shadow-graphs of the vorticity (top panel) and the
density (lower panel) field at the non-linear stage for the first Holmboe instability.}
\end{figure}

The evolution of the kinetic energy of the first Holmboe mode is shown in the top panel of 
figure \ref{fig8} with the solid line marked as H1.
The evolution of the two energies has similar behavior with the single mode investigations
at early times.
At times larger than $6<\gamma t$ there is an increase 
in the amplitude of kinetic and potential energy as well as an increase in the amplitude of the oscillations.
This behavior is due to the coupling of the different unstable Holmboe waves.
As shown 
in the the shadow-graphs 
of the vorticity (top panel) and density (bottom panel) of the first Holmboe mode
in figure \ref{fig9}
two of the initially four vortices that had developed in the linear stage have merged 
leading to three vortexes below the interface and four above the interface.

The evolution of the second Holmboe mode has also some differences than the single mode perturbation that were
previously examined. First, for this run it the mode with $k=3/4$ that dominates the non-linear behavior
and not the $k=1$ as in the single mode case examined in the previous section. 
The growth rates for the two wave numbers are very similar 
($\gamma=0.0156$ for $k=1$ and $\gamma=0.0150$ for $k=3/4$) and although both wave numbers were present
at early times $\gamma t \sim 4$) the wave number $k=3/4$ dominated. 
It is also worth noting that the decrease of the kinetic energy after the first peek that was observed
in the single mode run is not observed here. The potential and kinetic energies however have only 
slightly larger values than the ones observed in the single mode investigations.
When compared with the first Holmboe mode the potential energy of the second Holmboe mode is larger.
  
\begin{figure}
\includegraphics[width=8cm]{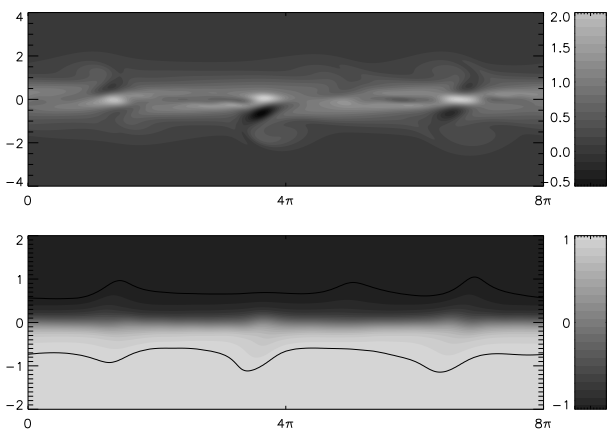}
\caption{\label{fig10} Shadow-graphs of the vorticity (top panel) and the
density (lower panel) field at the non-linear stage for the second Holmboe instability
with $\ell_{_X}=8\pi$.}
\end{figure}

Finally it should be observed that there is not a clear saturated 
stage of the potential energy observed in the simulations. 
This implies that the enchantment of vertical mixing of mass by fluid motion 
has not yet been fully suppressed by the nonlinearities.


\section{conclusions}

In this work the linear and non-linear evolution of stratified shear flow instabilities
for three different values of the Richardson number that span two orders of magnitude
was investigated.
The
three examined values of the Richardson number correspond  
to three different unstable modes 
namely the Kelvin Helmholtz mode for $Ri=0.2$ 
the first Holmboe mode for $Ri=2.0$ and  
the second Holmboe mode for $Ri=20$. 
All flows had identical velocity and density profiles so essential 
the only parameter changed was the amplitude of gravitational acceleration.

The linear investigation of the problem has shown that the inclusion of the viscosity and
diffusivity has strongly suppressed the region of instability for the two Holmboe modes
that extend to arbitrary large values of the Richardson number for the inviscid problem. 
For the examined value of the Reynolds number Re=500 the first Holmboe mode appears only
for $Ri<7$. In the range $13<Ri<30$ only the second Holmboe mode is present with
growth rate only three times smaller than the growth rate of the first mode
at an order of magnitude smaller $Ri$.
For strongly stratified environments therefore the higher Holmboe modes are 
the only modes that are able to destabilize the flow and generate turbulence
at finite $Re$.

The nonlinear development of the  three cases lead 
stretching of the density interface and enhanced mixing,
however not at the same level.
From the two Holmboe modes the first mode
resulted in a larger amplitude of kinetic energy
in the nonlinear stage, however the resulting potential of the second mode
exceeded that of first mode.
The Kelvin Helmholtz mode was the most dominant with the kinetic energy
of the perturbation reaching amplitudes hundred times bigger than that
and potential energy ten times bigger than that of the Holmboe modes in 
a much shorter times. However, if we take into account that the Richardson number  
has been increased by a factor of 100 a decrease of potential energy only by a factor of ten
is surprisingly small. Therefor there is non negligible amount of mixing even in very strongly stratified
environments. Thus in such flows turbulent mixing cannot be a priori excluded,
by virtue of the high value of the Richardson number.

Finally the issue of dimensionality is discussed. Two and three dimensional turbulence 
have a very different behavior the later having a much better efficiency at
generating small scales fast, dissipating  energy and diffusing advected scalars.
At the same time the cascade of energy will also enhance the dissipation of kinetic energy
and thus decrease the ability of the flow to convert it to potential energy.
It is therefor possible that the mixing behavior will be different in some aspects
than the results found here. These issues however are left to be investigated
in future work.


\begin{acknowledgments}

I would like to thank 
the Laboratoire de physique statistique at 
Ecole Normale Superieure and CNRS, for hosting 
and supporting me during the production of this work.
Some earlier computations were performed at 
the Observatoire de la Cote d'Azur
by IDRIS  CNRS  Grant No. 070597, and SIGAMM
mesocenter  OCA/University Nice-Sophia  and they are also thanked.
Finally,
I would also like to thank P.D. Mininni for his help while constructing 
the code.

\end{acknowledgments}

\appendix

\newpage
%

\end{document}